\documentclass[%
reprint,
superscriptaddress,
showpacs,showkeys,
 amsmath,amssymb,
 aps,
floatfix,
]{revtex4-1}

\usepackage{graphicx,subfigure}
\graphicspath{{Figures/}}
\usepackage{multirow}

\newcounter{reaction} 
\begin{document}
\draft


\title[]{Application of an extended van der Pauw method to anisotropic magnetoresistance measurements of ferromagnetic films}

\author{Movaffaq Kateb}
\affiliation{Science Institute, University of Iceland, Dunhaga 3, IS-107 Reykjavik, Iceland}

\author{Egill Jacobsen}
\affiliation{Science Institute, University of Iceland, Dunhaga 3, IS-107 Reykjavik, Iceland}

\affiliation{Science Institute, University of Iceland, Dunhagi 3, IS-107, Reykjavik, Iceland}

\author{Snorri Ingvarsson}
\email[Corresponding author email address: ]{sthi@hi.is}
\affiliation{Science Institute, University of Iceland,
Dunhaga 3, IS-107 Reykjavik, Iceland}

\begin{abstract}
We demonstrate anisotropic resistivity measurements using the extended van der Pauw (vdP) method in ferromagnetic Ni$_{80}$Fe$_{20}$ (Py) films. We apply it to measure anisotropic magnetoresistance (AMR) and compare the results of the vdP method with the more conventional  Hall-bar method along the hard and easy axis of the film and show that the vdP method gives more reliable AMR result. For instance the AMR result along the hard and easy axis of the film are in close agreement. 
Further, we applied the vdP method to study AMR in a series of Py films with thicknesses ranging between 10 -- 250~nm. The films were grown by sputtering deposition at an angle with respect to the substrate normal and with an \emph{in-situ} magnetic field, both conditions assisting in the definition of in-plane uniaxial anisotropy. The microstructure of Py films was characterized using X-ray reflectivity, diffraction and polar mapping of (111) planes. We detected no off-normal texture and negligible surface roughness, which indicates that self-shadowing is not dominating in our growth. Yet the films have well defined uniaxial anisotropy. Abrupt changes in the average resistivity vs.\ film thickness were observed, which cannot be explained by the models accounting for the thickness and grain size but strongly correlate with the changes in (111) texture in the films. We compared our results with the literature and show that independent of growth method, substrate and deposition temperature, the AMR value presents a saturation behavior with thickness at about 100~nm.
\end{abstract}

\maketitle

\section{Introduction}
Anisotropic magnetoresistance (AMR) is defined as the variation of resistance upon changing the relative alignment of electric current and magnetization \cite{mcguire1975}. Results of AMR measurements can be affected by the measurement technique as well as the sample preparation method \cite{mcguire1975}. For instance, Chikazumi \cite{chikazumi1950} measured AMR by sweeping the magnetic field while performing resistance measurement along the Ni$_{78.5}$Fe$_{21.5}$ bars. He showed that different bars magnetized in different directions present different AMR results. Bozorth \cite{bozorth1946} has shown that determining AMR by applying saturating fields parallel and perpendicular to current direction avoids erratic results in the early literature caused by ignoring the initial magnetization state of the specimen.

It was suggested already by McGuire and Potter \cite{mcguire1975} that use of van der Pauw (vdP) method might improve the precision in determining AMR properties of thin films. The vdP method is a simple and flexible technique to probe resistivity of uniform, continuous thin films of arbitrary shape \cite{vdpauw1958,Pauw1958}. However, the original vdP formalism is limited to specimens with isotropic resistivity ($\rho_{\rm{iso}}$). It has been shown that the vdP method can be extended to determine anisotropic resistivity \cite{Price1973} and that $\rho_{\rm{iso}}$ measured by the original technique is the geometric mean of resistivities along the principle axes of the film $\rho_{\rm{iso}}=\sqrt{\rho_1\rho_2}$ \cite{Pauw1958,Price1973}.

AMR results also depend on the thin film deposition technique, and in particular on the resulting microstructure and magnetization \cite{mcguire1975}. Control over magnetization direction can be achieved by applying \emph{in-situ} magnetic field during growth or by depositing under an angle with respect to the substrate normal \cite{Smith1959}. The origin of magnetic field induced anisotropy in permalloy Ni$_{80}$Fe$_{20}$ (Py), in which the effect of magnetostriction is negligible, was mainly attributed to directional ordering of Fe/Ni atom pairs \cite{chikazumi1950,Cullity2011}. The tilt angle, however, has been thought to induce anisotropy due to self-shadowing effect \cite{Smith1959}. Since then, it has been shown that self-shadowing leads to off-normal fibrous texture in Py \cite{Sun2007,solovev2017}. This encourages perpendicular (out-of-plane) anisotropy in the film and thus lowers in-plane anisotropy or even leads to in-plane isotropy, i.e\ a complete loss of in-plane anisotropy \cite{Zou2002}. However, it is still unclear how tilt deposition contributes to the in-plane uniaxial anisotropy.

There are a limited number of studies on the simultaneous utilization of \emph{in-situ} magnetic field and tilt deposition \cite{Sun2007,phuoc2013,Oliveira2014,Kateb2017}. Sun \emph{et al.} \cite{Sun2007} reported a deterioration of uniaxial anisotropy and loss of magnetic softness in CoZrO films by increase in tilt angle (0--55$^\circ$) in the presence of an \emph{in-situ} field of 400~Oe which assisted the easy axis induced by the tilt angle. Phuoc \emph{et al.} \cite{phuoc2013} showed that in the presence of a 200~Oe assisting magnetic field, the anisotropy field of Py/IrMn bilayer increases with increase in tilt angle, which is more pronounced for angles larger than 35$^\circ$. Oliveira \emph{et al.} \cite{Oliveira2014} studied 150~Oe competing field, i.e.\ a field perpendicular to the easy axis defined by the tilt angle, in the Cu/IrMn/Py/Cu system. They showed that the magnetic axis can be rotated $\sim$10--30$^\circ$ with respect to original easy axis induced by the tilt angle depending on the tilt angle. More recently, we studied deposition of Py under a 35$^\circ$ tilt angle with three different field configurations: zero field, assisting and competing \emph{in-situ} saturation field of 70~Oe \cite{Kateb2017}. In our study we showed that tilt angle always determines the easy axis, even if the applied field competes with the easy axis induced by the tilt angle. It was also shown that a combination of tilt angle and assisting field results in very well defined uniaxial anisotropy in the Py i.e. square easy axis with sharp switching and linear hard axis without hysteresis.

In this work, we use the extended vdP method for AMR measurements in Py films and compare it with the more conventional method of defining Hall-bar patterns in the films. We use the method to study a series of different thickness Py films, prepared by tilt deposition with an assisting \emph{in-situ} magnetic field, to make sure they present well defined uniaxial anisotropy. We do careful X-ray measurments in studying the microstructure, e.g.\ texture, of our films and compare our results with the literature. We find that there is a change in microstructure as the films increase in thickness, and that this is reflected in the resistivity measurements.
\section{Experimental method}
\subsection{Magnetoresistance measurements}
\begin{figure}
    \centering
    \includegraphics[width=1\linewidth]{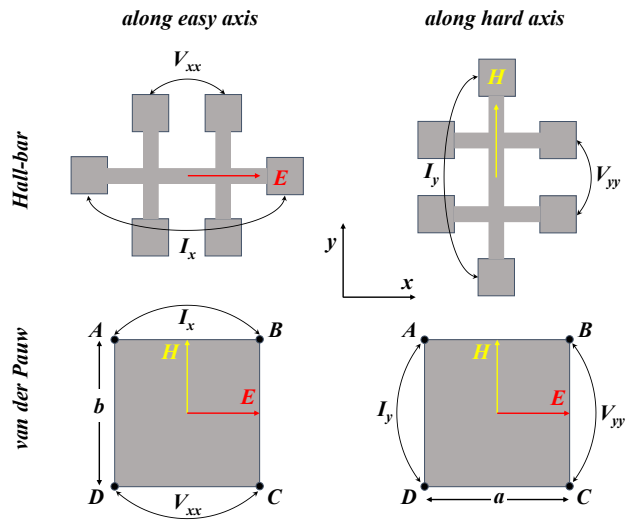}
    \caption{Schematic illustration of resistance measurements along hard (H) and easy (E) axis for both hall-bar and vdP methods.}
    \label{fig_scheme1}
\end{figure}

We compared the magnetoresistance obtained by vdP and by lithographically patterned Hall-bars. To this end, clean (001) p-Si was dehydrated at 140~$^\circ$C for 5~min on a hotplate and then exposed to HMDS vapor for 5~min to become more hydrophilic. Then 1--2~mL maN-1410 photoresist (Micro resist tech. GmbH) was dispensed and spin coated at 4000~rpm for 300~s (pre-spin at 500~rpm for 15~s) and soft baked at 100~$^\circ$C for 90~s. This gives 1~$\mu$m thick resist on a 4" wafer. After 6~s exposure (DUV-1000 AB-M Inc. mask aligner) at 25~mW/cm$^2$ the pattern was developed in maD-533/S (Micro resist tech. GmbH) for 30~s and rinsed with DI water and dried with N$_2$. Square (15$\times$15~mm$^2$) and Hall-bar (0.4$\times$1.6~mm$^2$) patterns were grown simultaneously and prepared with a lift-off in Acetone. The growth process included a 4~nm thick Cr underlayer, for adhesion, and a 40~nm thick Py film. During deposition, an \emph{in-situ} magnetic field of 70~Oe was applied to induce uniaxial anisotropy in the desired direction, without the aid of the tilt angle. The Hall-bars were made large enough that the in-plane shape anisotropy of the Hall-bar structure would not affect the magnetization direction induced during growth. Then the AMR was measured as shown schematically in figure \ref{fig_scheme1} by driving current through the Hall-bar and measuring voltage at the side contacts.
%

The vdP method is a useful technique to probe resistivity of uniform, continuous thin films of arbitrary shape \cite{vdpauw1958,Pauw1958}. In the vdP method, four small contacts must be placed on the sample perimeter. We choose to work with square samples, with electrical contacts at each of the four corners labeled $A$, $B$, $C$, and $D$ as illustrated in Fig.~\ref{fig_scheme1}. While the resistivity in magnetic materials is clearly anisotropic, the original vdP method assumes the film is isotropic. The corresponding isotropic resistivity value, $\rho_{\rm{iso}}$ is obtained by:
\begin{equation}
	\exp\left(-\frac{\pi d}{\rho_{\rm{iso}}}R_{AB,CD}\right)+\exp\left(-\frac{\pi
	d}{\rho_{\rm{iso}}}R_{AD,CB}\right)=1
\end{equation}
where $d$ is the film thickness and e.g.\ $R_{AB,CD}$ is the resistance obtained by applying current to $AB$ and picking up the voltage at the opposite side
between $CD$ or \emph{vice versa}. It has been shown that $\rho_{\rm{iso}}=\sqrt{\rho_x \rho_y}$ is the geometric mean of principle resistivities in anisotropic thin films \cite{Pauw1958,Price1973} i.e. along the easy ($\rho_x$) and the hard ($\rho_y$) axes in our case. The ratio of principle resistivities can be obtained from Price's \cite{Price1973} extension to the vdP method for anisotropic samples, as below:
\begin{equation}
    \sqrt{\frac{\rho_{x}}{\rho_{y}}}=-\frac{b}{\pi a}\ln\left(\tanh \left[\frac{\pi
    dR_{AD,CB}}{16\rho_{\rm{iso}}}\right]\right)
    \label{eq:rhoratio}
\end{equation}
where $b$ and $a$ are the side lengths of a rectangular sample and $R_{AD,BC}$ is resistance along the $b$ sides as described above. Eq.~(\ref{eq:rhoratio}) yields the ratio of easy and hard axis resistivity. The individual values of principle resistivities can subsequently be obtained by:
\begin{equation}
	\rho_{x}=\rho_{\rm{iso}}\sqrt{\frac{\rho_{x}}{\rho_{y}}}
    \label{eq:x}
\end{equation}
\begin{equation}
	\rho_{y}=\rho_{\rm{iso}}\left(\!\sqrt{\frac{\rho_{x}}{\rho_{y}}}\;\right)^{-1} \quad
    \label{eq:y}
\end{equation}

Our magnetoresistance measurements were done at in-plane saturation field of $\sim$23~Oe parallel and perpendicular to the current direction, respectively. The strength of the field is enough to saturate the magnetization as it is 10 times the coercive field $H_c$ and 5 times of $H_k$ in our thinnest films. All measurements were performed at room temperature. Since high current densities may perturb local magnetization \cite{Li2004} and/or produce heating, care was taken to use low current densities in resistivity measurements. Thus we swept between $\pm$10~mA for measuring vdP and Hall-bar resistivities (the I-V curves were perfectly linear within this range). 
The AMR ratio is given by \cite{mcguire1975}:
\begin{equation}
	\rm{AMR}=\frac{\Delta\rho}{\rho_{\rm{ave}}}=\frac{\rho_{\|}-\rho_{\bot}}{\rho_{\rm{ave}}}
	\label{eq:amr}
\end{equation}
where $\rho_{\rm{ave}}$ for FCC materials like Py defined as \cite{mcguire1975}:
\begin{equation}
	\rho_{\rm{ave}}=\frac{1}{3}\rho_{\|}+\frac{2}{3}\rho_{\bot}
\end{equation}
where $\rho_{\|}$ and $\rho_{\bot}$, respectively, are resistivities with magnetization saturation parallel and perpendicular to the current direction. Thus, each of the $\rho_{x}$ and $\rho_{y}$ can be translated to $\rho_{\|}$ and $\rho_{\bot}$ by applying proper external magnetic field. To this end three vdP measurement was performed i.e.\ without applied field ($B_0$), with saturation field along the easy ($B_{\rm easy}$) and hard ($B_{\rm hard}$) axes, respectively. This yields values for $\rho_{\rm{iso}}$ at $B_0$, $B_{\rm easy}$ and $B_{\rm hard}$ that can be translated to $\rho_{\rm easy}$ and $\rho_{\rm hard}$ using Eq. (\ref{eq:x}) and (\ref{eq:y}). Now, for example, AMR along easy axis is determined by considering $\rho_{\|}$ to be $\rho_{\rm easy}$ at $B_{\rm easy}$ and $\rho_{\bot}$ equal to $\rho_{\rm easy}$ at $B_{\rm hard}$. In a similar manner one can define AMR along the hard axis by using $\rho_{\|}=\rho_{\rm hard}$ at $B_{\rm hard}$, and $\rho_{\bot}=\rho_{\rm hard}$ at $B_{\rm easy}$.

The definition of AMR does not contain any reference to sample geometry or its crystalline or other anisotropies, only the angle between magnetization and current direction. Assuming the $x'$-axis being some general current direction, one can determine $\rho_{x'}$:
\begin{equation}
    \rho_{x'}=\rho_{\|}+\Delta\rho cos^2\theta
    \label{eq:rhoxx}
\end{equation}
here $\theta$ stands for angle between current ($x'$) and saturated magnetization direction. It is worth noting that Eq.~(\ref{eq:rhoxx}) states the resistivity is only dependent on $\theta$ and not on the zero field easy and hard axis directions.

\subsection{Thickness series grown by field assisted tilt sputtering}
\begin{figure}
    \centering
    \includegraphics[width=1\linewidth]{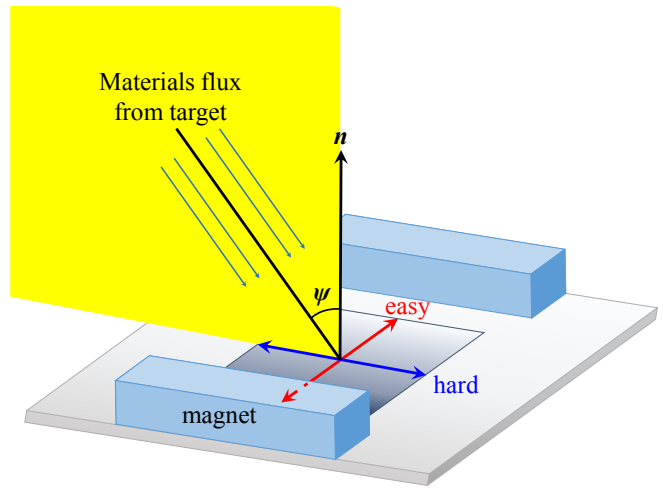}
    \caption{Schematic illustration of deposition geometry and tilt angle induced hard and easy axis in the film.}
    \label{fig_scheme2}
\end{figure}

Our deposition configuration for studying the effect of different thickness is shown schematically in figure \ref{fig_scheme2}. Regardless of growth method, e.g.\ evaporation \cite{Smith1959}, DC or RF magnetron sputtering \cite{Li2015} or even presence/absence of applied field \cite{Kateb2017}, tilt deposition has been found to determine the magnetic easy axis in Py.

Our series of samples with different thickness was grown on (001) p-Si with a 100~nm thick layer of thermally grown oxide. No underlayer was used since we have found that tilt deposited underlayer increases anisotropy field in our films, as has been reported by others \cite{Sun2007,Li2015, McMichael2000}. Our depositions were carried out in a UHV ($<5\times10^{-9}$~mbar base pressure) magnetron sputter system at a pressure of $1.3\times 10^{-3}$~mbar and 150~W which results in 1.20~\r{A}/s deposition rate. The deposition angle was 35$^\circ$ with respect to the substrate normal, with a target to substrate distance of 20~cm. During deposition, a magnetic field of 70~Oe was applied using a pair of permanent magnets attached to the sample holder. The entire sample holder rotated around the substrate normal $n$ 360$^\circ$ back and forth at $\sim$12.8~rpm. The process of stopping and reversing takes 200~ms. The rotation is necessary in order to obtain uniform film thickness, while the stop time before reversal is what determines the magnetization axis along with the tilt angle. Thickness uniformity over large area was examined simply by lifting-off pre-patterned lines (from side to side and along diagonals of our $20\times20$~mm$^2$ substrates) followed by step height measurement using atomic force microscopy (AFM).

X-ray diffraction (XRD) was carried out using a X'pert PRO PANalitical diffractometer (Cu K$_\alpha$ line, wavelength 0.15406~nm) mounted with a hybrid monochromator/mirror on the incident side and a 0.27$^\circ$ collimator on the diffracted side. A line focus was used with a beam width of approximately 1~mm. Grazing incidence (GI)XRD scans were carried out with the incident beam at $\theta = 1^\circ$ with 0.05$^\circ$. The film thickness, density and surface roughness was determined by low-angle X-ray reflectivity (XRR) measurements with 0.005$^\circ$ angular resolution. XRR measurements were fitted using a commercial X'pert reflectivity program based on Parrat formalism \cite{parratt54:359}. Further, pole scans were done which enable detecting the texture evolution in tilt sputtered films \cite{Mahieu2006}. Briefly, a pole scan is done for a specific d-spacing, i.e.\ a fixed $\theta-2\theta$ peak while the specimen is rotated in-plane ($\phi$) at different out-of-plane ($\psi$) angles. Normally, a single pole scan is not enough to fully determine the orientation distribution within a specimen. However, since our films are polycrystalline the main focus of the present study is on the pole figure for (111) planes. We measured the (111) pole figures by setting $\theta-2\theta$ to the corresponding peak obtained in the normal XRD. We also repeated such measurement for the bare substrate and found that the raw pole figures of the films were affected by the substrate pattern. Thus we subtracted the substrate pattern from raw pole figures assuming they are collected at identical conditions. To emphasize the changes in $\langle111\rangle$ texture with the film thickness we subtracted the pole figure of each film by the film with the next lower thickness, e.g.\ we subtracted the pole figure of 40~nm from 50~nm, 50~nm from 75~nm and so on. Thus we might be able to detect texture evolution more clearly.

To obtain hysteresis loops, we used a high sensitivity magneto optical Kerr effect (MOKE) looper. The detail on the MOKE setup can be found in reference \cite[Sec.~3.2.1]{jacobsen2016}. We used a constant field step of 0.2~Oe along the easy axis and 1~Oe along the hard axis of the film in the range $\pm$60~Oe, with dwell time of 300~ms. Coercive field ($H_c$) was read directly from the loop widths along the easy axis and anisotropy field was obtained by extrapolating the hard axis MH curve at low field to high field values and determining the field for which the material saturates. It is worth mentioning that the latter is a common method in the case when there are well defined easy and hard directions at right angles.

For the thickness series all resistivity and AMR measurements were done using vdP method.
\section{Results and discussions}
\subsection{Magnetoresistance measurements}
We compared the result of vdP with the Hall-bar method for identical test samples of Py, deposited simultaneously and with the sputter at normal incidence. Table \ref{tab:HbvsvdP} summarizes the result of resistivity measurements by both methods at different magnetization i.e. $B_0$, $B_{\rm easy}$ and $B_{\rm hard}$. The table also contains primary magnetoresistance values of $\Delta\rho$, $\rho_{\rm{ave}}$ and AMR. The results show that the vdP method always presents lower resistivity values than the Hall-bar method. The difference in $\rho_{\rm{ave}}$ along the hard and easy axis is 7.7\% employing the Hall-bar method. This is noticeable error since the $\rho_{\rm{ave}}$ is measured at saturated magnetization and must be independent of the zero field magnetization state of the specimen \cite{bozorth1946}. While, the $\rho_{\rm{ave}}$ difference in the vdP method presents a more reasonable 1.7\% difference which is in better agreement with the definition of $\rho_{\rm{ave}}$. Finally, the AMR values determined by the vdP method show an absolute difference of 0.11\% along the easy and hard directions, while conventional Hall-bar method result in 0.5\% absolute difference in AMR.

\begin{table*}[]
    \centering
    \caption{Resistivity measurements using vdP and Hall-bar methods along hard and easy axes (according to figure \ref{fig_scheme1}) in
    $\mu\Omega\,$cm. The $\|$ and $\bot$ in the superscript are denoting $\rho_{\|}$ and $\rho_{\bot}$ in each direction.}
    \label{tab:HbvsvdP}
    \begin{tabular}{|c|c|c c c|c|c|c|}
    \hline\hline
	method & current & \multicolumn{3}{c|}{$\rho_{\rm{current-direction}}(B)$} & $\rho_{\|}-\rho_{\bot}$ &$\rho_{\rm{ave}}$ & AMR(\%) \\
    & direction & $B_{0}$ & $B_{\rm{easy}}$ & $B_{\rm{hard}}$ & & & \\
    \hline
    Hall-bar & easy & 33.6911 & 33.4460$^\|$ & 33.0269$^\bot$ & 0.41902 & 33.1666 & 1.2634 \\
    Hall-bar & hard & 35.8481 & 35.7178$^\bot$ & 36.3524$^\|$ & 0.6346 & 35.9293 & 1.7672 \\
    vdP & easy & 32.0508 & 32.1325$^\|$ & 31.4900$^\bot$ & 0.6425 & 31.9183 & 2.0130 \\
    vdP & hard & 30.9832 & 30.9697$^\bot$ & 31.5251$^\|$ & 0.5963 & 31.3673 & 1.9010 \\
    \hline\hline
    \end{tabular}
\end{table*}

It is worth noting that the Hall-bar method provides a very simple route for direct measurement of magnetoresistance, i.e.\ simply by recording resistance values when sweeping applied field parallel and perpendicular to the bar. The vdP method however, demands that one switches contacts during measurement and requires more, albeit rather simple, data processing. Besides $\rho_{\rm iso}$ obtained by original vdP equation does not show any change at $B_0$, $B_{\rm hard}$ and $B_{\rm easy}$ and had to be extended for AMR measurement. Having verified the extended vdP method for measuring AMR we apply it to tilt deposited samples of different thickness.

\subsection{Thickness series grown by field assisted tilt sputtering}
\subsubsection*{Growth calibration:}
In general, thickness uniformity is a major problem in tilt deposition. However, rotation of the substrate in our geometry (cf. figure \ref{fig_scheme2}) resolves this issue as confirmed by AFM measurement to be less than 3~{\AA} in a $18\times18$~mm$^2$ area for a 30~nm thick film. While for the film grown with identical conditions but without rotation, a 6~nm difference in thickness was determined.

\begin{figure}
    \centering
    \includegraphics[width=1\linewidth]{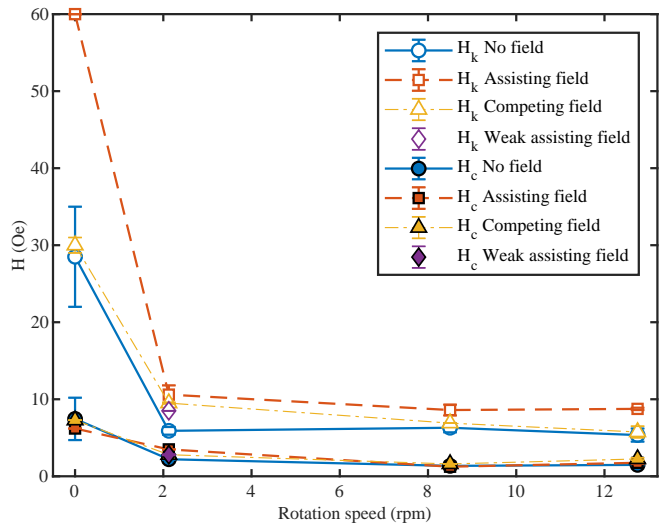}
    \caption{Effects of rotation speed on the $H_k$ (hollow symbols) and $H_c$ (filled symbols) of films deposited at a 35$^\circ$ angle with respect to the substrate normal} without field and with field assisting and competing tilt angle.
    \label{fig_Rotation}
\end{figure}

Figure \ref{fig_Rotation} shows the effects of rotation on the
$H_k$ and $H_c$ of the samples grown without field and
with field either assisting or competing with the tilt angle. The samples without rotation present very high values of both $H_k$ and $H_c$ which might be due to thickness non-uniformity.
However, low rotation speed of $\sim$2.1~rpm causes a sharp drop in all cases and results seem unchanged at higher speeds. It is worth noting the $H_k$ obtained in the assisting field configuration is always higher than the competing and no-field counterparts. Further we also tried applying a much weaker assisting field of 20~Oe which gives intermediate $H_k$ and $H_c$ between those of assisting field and no-field as shown by diamond markers in the figure. Thus, the stronger \emph{in-situ} field gives higher $H_k$ in our case.

\subsubsection*{X-ray reflectivity (XRR):}
\begin{figure}
    \centering
    \includegraphics[width=1\linewidth]{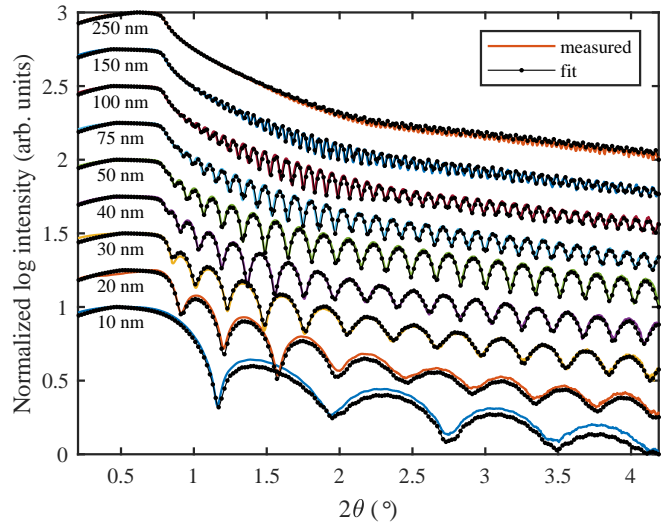}
    \caption{The measured and fit XRR curves of different Py films with 10--250~nm thicknesses. The curves shifted manually for clarity. Our films were deposited at a 35$^\circ$ angle with respect to the substrate normal with assisting \emph{in-situ} magnetic field of 70~Oe at $\sim$12.8~rpm rotation speed.}
    \label{fig_XRR}
\end{figure}

The thickness, density and roughness of the films were obtained by fitting the XRR curves shown in Fig.~\ref{fig_XRR}. The plots shown are raw data, for clarity we omit the fits, but they agree almost perfectly with the data. To get the best fits both PyO and water layers were included in the fit procedure, which is reasonable since the measurements were done in air atmosphere. The thicknesses of the samples indicate a stable growth rate of precisely 1.2~{\AA}/s. The densities of the films show very slight fluctuation about their average of 8.61~g/cm$^3$. The uncertainty in the fit results decreases at higher thicknesses since there is more film material exposed by X-ray and more fringes to fit. The surface of the films present excellent smoothness, with X-ray results showing about 2~{\AA} roughness. For a few samples the roughness was verified by AFM, which gave results in the range 1.9--2.8~{\AA} RMS. Even though there appeared to be a slight difference in the roughness obtained in XRR fit and AFM both method verifying extreme smoothness of our films. Since self-shadowing is associated with appearance of noticeable surface roughness, a negligible surface roughness obtained here indicates that self-shadowing is not dominating in our growth. Thus its contribution to the obtained uniaxial anisotropy is highly doubtful. 

\begin{table*}
	\centering
	\label{tab:XRRfit}
	\caption{The film thickness, density and surface roughness obtained by fitting of the XRR results.}
	\begin{tabular}{|l|c c c c c c c c c|}
    	\hline\hline
		sample&10&20&30&40&50&75&100&150&250\\
        \hline
Thickness (nm)&10.49&21.47&29.63&41.41&50.59&75.49&103.62&157.07&259.78\\
Density (g/cm$^3$)&8.41&8.65&8.65&8.58&8.63&8.63&8.63&8.63&8.65\\
Roughness ({\AA})&2&1.25&2&2.01&2&2.02&2&1.11&2\\
        \hline\hline
	\end{tabular}
\end{table*}

\subsubsection*{Microstructure}
\begin{figure}
    \centering
    \includegraphics[width=1\linewidth]{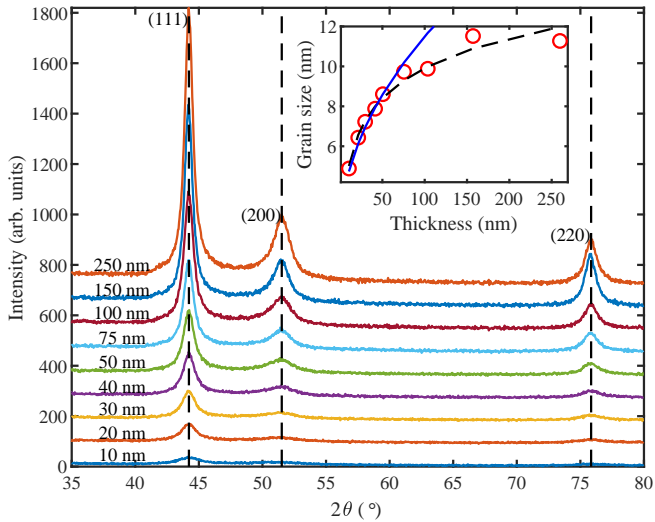}
    \caption{GIXRD pattern of Py with different thicknesses. The vertical dashed lines are indicating the bulk peak positions. The inset shows the estimated grain size (red circles) from the Scherrer formula compared with square-root (solid blue) and natural logarithm (dashed black)
    dependency on thickness. Our films were deposited at a 35$^\circ$ angle with respect to the substrate normal with assisting \emph{in-situ} magnetic field of 70~Oe at $\sim$12.8~rpm rotation speed.}
    \label{fig_GIXRD}
\end{figure}

Figure \ref{fig_GIXRD} shows the GIXRD pattern of the films which is a common technique in characterization of polycrystalline thin films. For the 10~nm thick sample the only detectable peak is (111) at 44.22$^{\circ}$ while all our thicker films also present (200) and (220) peaks at 51.75 and 76.02$^{\circ}$, respectively. The strong (111) peak followed by intermediate (200) and weak (220) is sometimes referred to as a characteristic of sputtered Py \cite{Dzhumaliev2016}. The increase in the (111) peak height with thickness indicates an enhancement in crystallinity which can be quantified using the Scherrer formula. The estimated grain size from the (111) peak is shown by red circles in the figure inset. Typically the grain size of FCC elements grows as the square root of the film thickness \cite{Neerinck1996}. In our case, however, at thicknesses above 50~nm the grain size deviates from a square root behavior (indicated by the solid blue line). This has also been observed by Neerinck~\emph{et al.} \cite{Neerinck1996}, investigated by transmission electron microscope (TEM) imaging, finding that grain boundary pinning by voids limits normal grain growth in sputtered Py films. They also showed that estimated grain size, by the Scherrer formula from GIXRD pattern, is in agreement with the grain size observed by TEM. The dashed line shows natural logarithm which is plotted to aid the eye. We emphasize that our 150~nm thick sample presents the largest estimated grain size. In addition, the (220) peak of this sample is higher than the rest of the films with respect to the background intensity. This indicate a variation of growth mode at different thickness.

We emphasize that the incident angle is fixed in GIXRD, at 1$^\circ$ in our case, and thus the normal vector of detected planes rotates during measurement. (The only detection angle where it is normal to the substrate would be at $2\theta=2^\circ$). This means that the (111), (200) and (220) peaks are obtained at different angles with respect to the substrate. This should be kept in mind when comparing them.

\begin{figure}
    \centering
    \includegraphics[width=1\linewidth]{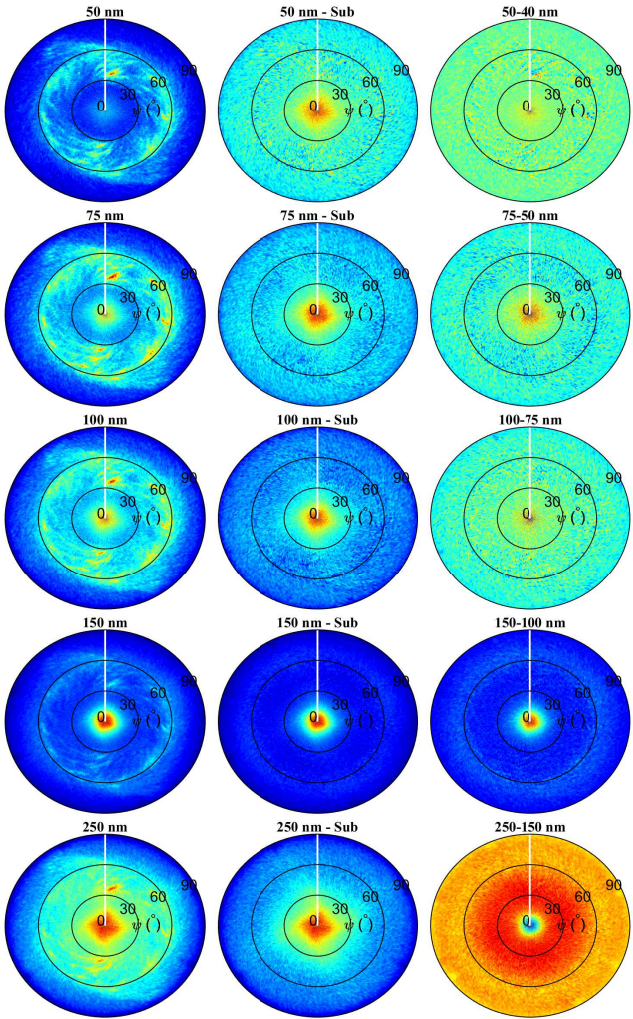}
    \caption{Normalized intensity pole figures of the (111) peak for 50--250~nm samples: unmodified pole figure (left), after subtracting the pure substrate pattern (middle), and after subtracting pole figure from the next thinner film (right). The white line along the north indicates the direction of the sputtered flux during 200~ms stop-and-turn. Our films were deposited at a 35$^\circ$ angle with respect to substrate normal with an assisting \emph{in-situ} magnetic field of 70~Oe at $\sim$12.8~rpm rotation speed.}
    \label{fig_Pole}
\end{figure}

Figure \ref{fig_Pole} shows the (111) pole figures of Py films with thicknesses higher than 50~nm. It has been shown that determining the texture of polycrystalline Py with thickness below 35~nm is rather difficult \cite{Rijks1997}. The left column shows the original pole figures which has a vortex like pattern between $\psi=35-65^{\circ}$. It originates from the substrate, due to the interaction depth of the X-ray ($\sim$2~$\mu$m) compared to small thickness of the films. In an attempt to remove the substrate pattern we subtracted a pure substrate pole figure from the films' pole figures, assuming the pole figures of the films are obtained under reasonably identical conditions as the substrate.

The pole figures after subtracting the substrate pattern become more identical to each other as shown in the middle column of figure \ref{fig_Pole}. All these pole figures show an intense spot in the center indicating $\langle$111$\rangle$ is preferred growth direction normal to the substrate ($\psi=0$) without any tilted texture. The $\langle$111$\rangle$ fiber texture normal to the substrate has been detected previously in Py using TEM \cite{Katada2002}. It has also been shown that the rotation of the sample during growth encourages texture normal to the substrate \cite{Li2004}, rather than a tilted texture. Again this indicates the self-shadowing is not dominant here and its contribution to the obtained uniaxial anisotropy is doubted. It can be seen that the sharpness of the $\langle$111$\rangle$ spot is increasing from ``50~nm -- Sub" to ``150~nm -- Sub". The full width at half maximum (FWHM) is $\sim$12$^{\circ}$ for ``150~nm -- Sub". However, for ``250~nm -- Sub" the (111) spot is considerably broader with a FWHM of $\sim$40$^{\circ}$. This indicates the contribution of a different mechanism to growth between thickness of 150 and 250~nm. To investigate this further we subtracted each pattern by one step lower in thickness, e.g.\ 250 -- 150~nm. This enables understanding the dominant growth mechanism or transitions assuming identical growth conditions for samples. This \emph{incremental subtraction} is shown in the right column of figure \ref{fig_Pole}. The 50 -- 40~nm and 100 -- 75~nm show a scatter pattern similar to ones obtained for our polycrystalline Py target material (not shown here), but with slightly higher intensity at the center. However, the 75 -- 50~nm pattern shows a weaker scatter pattern and a very intense $\langle$111$\rangle$ spot at the center. This variation indicates a competition between the $\langle$111$\rangle$ perpendicular texture and equiaxed grain growth. In the 150 -- 100~nm the pattern consist of an intense $\langle$111$\rangle$ spot at the center indicating the $\langle$111$\rangle$ fiber texture becoming completely dominant. There is also a faint ring at $\sim$70$^{\circ}$ which belongs to $\langle$11$\overline{1}\,\!\rangle$ and is consistent with the $\langle$111$\rangle$ being perpendicular to the substrate. The 250 -- 150~nm presents a more intense ring between 10--50$^{\circ}$ compared to the spot at the center. This might be due to knock-on deformation \cite{Mahieu2006}.
\subsubsection*{Magnetic properties:}
\begin{figure}
    \centering
    \includegraphics[width=1\linewidth]{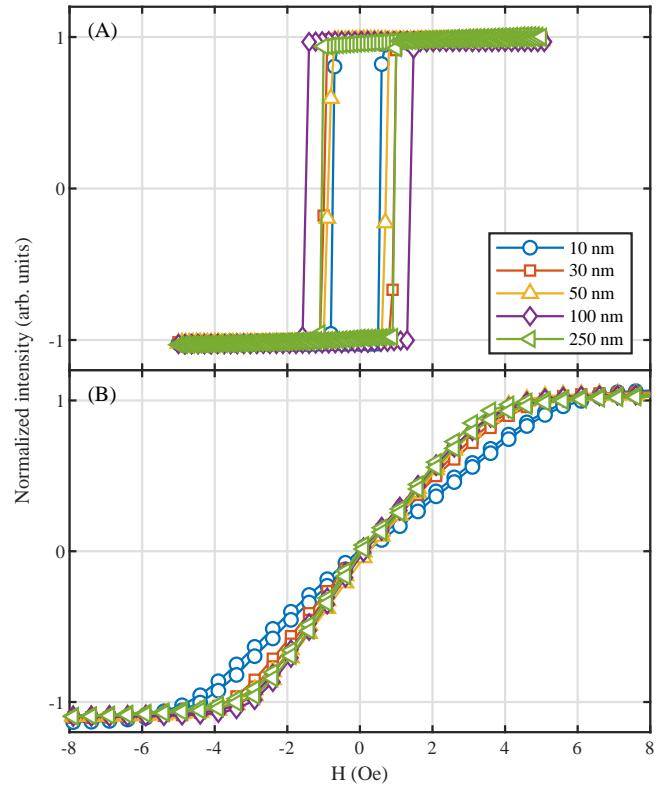}   
    \caption{MOKE hysteresis loops along the (A) easy axis and (B) hard axis of the films with various thicknesses. Our films were deposited at a 35$^\circ$ angle with respect to the substrate normal with assisting \emph{in-situ} magnetic field of 70~Oe at $\sim$12.8~rpm rotation speed.}
    \label{fig_MOKE}
\end{figure}

Figure \ref{fig_MOKE} shows the hysteresis loops along the both easy and hard axis of the film induced by field assisted tilt deposition. It can be seen that for a wide range of thicknesses this method gives very well defined uniaxial anisotropy i.e.\ square easy axis with sharp switching and linear hard axis without hysteresis. The $H_c$ shows no systematic change with thickness or grain size of the film. However, the 100~nm thick film presents a maximum $H_c$ of 1.4~Oe. Miyazaki \emph{et al.} \cite{Miyazaki1989} reported such a maximum and explained it by the change in domain wall structure. For the case of Py it changes from N{\'e}el to crosstie and then to Bloch walls at about 50--60 and 90--100~nm, respectively \cite{Ahn1966,Freedman1969}.
Along the hard axis, maximum $H_k$ of 5.4~Oe was obtained for 10~nm thik film. As the film thickness increases, $H_k$ drops and gives constant value of $\sim$3.4~Oe for thicknesses higher than 50~nm.

\subsubsection*{Magnetoresistance:}
\begin{figure}
    \centering
    \includegraphics[width=1\linewidth]{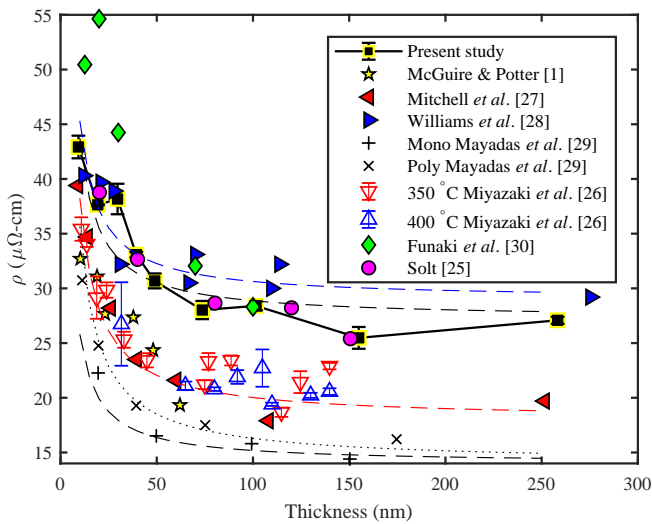}
    \caption{Variation resistivity with thickness in present study in comparison with previous results. The solid black line in our result is plotted to aid the eye. The dashed lines indicate fitting to Fuchs model to our result and to that of Mitchel \emph{et al.} \cite{Mitchell1964}, Williams \emph{et al.} \cite{Williams1986} and single crystal films of Mayadas \emph{et al.} \cite{Mayadas1974} while the dotted line indicates fitting Mayadas-Shatzkes models to the annealed polycrystalline films of Mayadas \emph{et al.} \cite{Mayadas1974}. Our films were deposited at a 35$^\circ$ angle with respect to the substrate normal with assisting \emph{in-situ} magnetic field of 70~Oe at $\sim$12.8~rpm rotation speed.}
    \label{fig_rho}
\end{figure}

Figure \ref{fig_rho} shows the dependency of $\rho_{\rm{ave}}$ on the film thicknesses in comparison with other studies in the literature. These results are collected using different methods in a long time span and are obtained from samples with different preparation as summarized in Table~\ref{tab:Previous}. It can be seen that we used a similar method, deposition rate (1.20~{\AA}) and substrate to Solt \cite{Solt1985}. As a result the resistivity values obtained in both experiments are in close agreement. It is worth mentioning that the samples in Ref.~\cite{Solt1985} were annealed for a few hours at 250$^\circ$, which according to Miyazaki \emph{et al.} \cite{Miyazaki1989} has negligible effect on the microstructure. A similar trend and close agreement can be found in the resistivities of Mitchell \emph{et al.} \cite{Mitchell1964} and Miyazaki \emph{et al.} \cite{Miyazaki1989}. Both used evaporation with nearly identical substrate temperature and deposition rate.

\begin{table*}[]
    \centering
    \caption{Sample preparation condition in previous studies where T, H and t denote temperature, applied field and anneal time, respectively..}
    \label{tab:Previous}
    \footnotesize{
    \begin{tabular}{|r|c c c c|c c c|c|c|}
    \hline\hline
    \multicolumn {1}{|c}{} & \multicolumn {4}{|c|}{Growth} & \multicolumn {3}{c|}{Anneal} & \multicolumn {1}{c|}{$\rho_{0}$} & Ref. \\
    & Rate (\r{A}/s) & substrate & T ($^\circ$C) & H (Oe) & T ($^\circ$C) & H (Oe) & t
    (h) & $(\mu\Omega\,$cm) & \\
    \hline
    \parbox[t]{2mm}{\multirow{5}{*}{\rotatebox[origin=c]{90}{Evaporation}}}
    & 16 & glass & 300 & 20 & - & - & - & 18 & \cite{Mitchell1964} \\
    & 1000 & glass & 300 & 1200 & - & - & - & 28 & \cite{Williams1986} \\
    & 2.3 & MgO & 300 & 60 & - & - & - & 14 & \cite{Mayadas1974} \\
    & 2.3 & MgO & 25 & 60 & 300 & - & 2 & 14 & \cite{Mayadas1974} \\
    & 10 & glass & 350-400 & 25 & - & - & - & 18 & \cite{Miyazaki1989} \\
    \hline
    \parbox[t]{2mm}{\multirow{3}{*}{\rotatebox[origin=c]{90}{Sputter}}}
    & 5 & glass & 25 & 0 & 150-450 & 500 & 1 & 27 & \cite{Funaki1994} \\
    & - & - & - & - & 400 & - & 1 & 14.5 & \cite{mcguire1975} \\
    & 1.16 & SiO$_{2}$-Si & 25 & 400 & 250 & 400 & 3 & 24 & \cite{Solt1985} \\
    \hline\hline
    \end{tabular}
    }
\end{table*}

Furthermore an approximation of Fuchs' theory \cite{Fuchs1938} (Eq. ~(\ref{eq:Fuchs})) plotted in the figure with dashed lines.
\begin{equation}
    \rho=\rho_{0}\left(1+\frac{3}{8}\frac{\lambda_{0}}{d}\right)
    \label{eq:Fuchs}
\end{equation}
where $\rho_{0}$ and $\lambda_{0}$ are the bulk resistivity and mean free path of electrons, respectively. Mayadas \emph{et al.} \cite{Mayadas1974} pointed out $\rho_{0}\lambda_{0}$ should be constant, e.g.\ equal to 31.5$\times10^{-6}~\mu\Omega\,\rm cm^2$ for Py, thus independent variation of parameters is restricted. The best fits for our films give $\rho_{0}$ of 27 $\mu\Omega\,$cm considerably higher than the bulk value of 14.5 $\mu\Omega\,$cm reported by Bozorth \cite{Bozorth1951}. While results of single crystal films by Mayadas \emph{et al.} \cite{Mayadas1974} present a reasonable value of 14~$\mu\Omega\,$cm for the bulk. This difference indicates that the simple surface scattering assumption in the Fuchs' model is not satisfied in polycrystalline films. 

For polycrystalline films, Mayadas and Shatzkes \cite{Mayadas1970} developed the following model to take into account grain boundary scattering:
\begin{equation}
    \frac{\rho_{0}}{\rho}=3\left[\frac{1}{3}-\frac{\alpha}{2}+\alpha^{2}-\alpha^{3}ln\left(1+\frac{1}{\alpha}\right)\right]
    \label{eq:MS}
\end{equation}
\begin{equation}
    \alpha=\frac{\lambda_{0}}{d}\frac{\Re}{1-\Re}
    \label{eq:alpha}
\end{equation}
where $\Re$ is the coefficient of grain boundary reflection. 

The Mayadas-Shatzkes model is also plotted in figure \ref{fig_rho} using a dotted line, which perfectly fits $\rho$ values of polycrystalline films reported by Mayadas \emph{et al.} \cite{Mayadas1974}. However, they annealed their samples in a way that the grain size is equal to or bigger than the film thickness. They also mentioned the as deposited polycrystalline film cannot be fitted with this model since $\lambda_{0}$ might change drastically with the thickness. This issue is also limiting for ultra-thin films when contribution of surface and grain boundary scattering to excess resistivity is the same \cite{Mayadas1974}.

It can be seen in figure \ref{fig_rho} there are some abrupt changes in the all data sets (e.g.\ 30, 70 and 150~nm in our result and 40~nm for McGuire and Potter) which cannot be explained by surface and grain boundary scattering. We found strong correlation between growth of $\langle111\rangle$ texture and deviation of $\rho_{\rm iso}$ from Fuchs' model. For instance 75 -- 50 and 150 -- 100~nm in which growth of $\langle111\rangle$ texture is dominant (cf. figure \ref{fig_Pole} right column) present more deviation from Fuchs' model and lower resistivity. 

\begin{figure}
    \centering
    \includegraphics[width=1\linewidth]{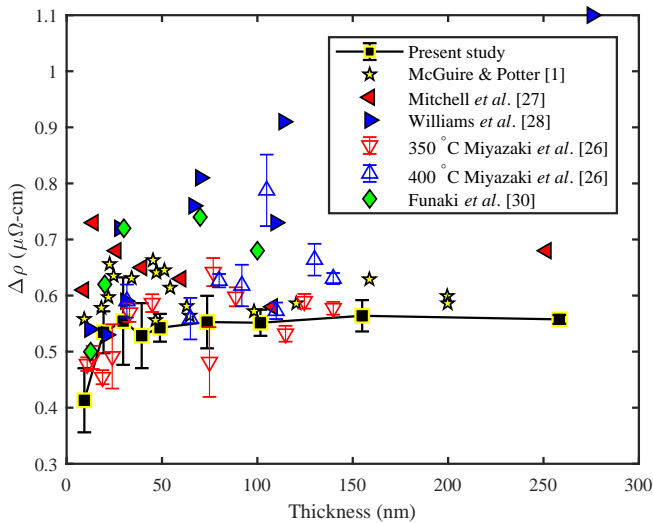}
    \caption{The change in $\Delta\rho$ with the film thickness in comparison with previous results. The solid black line in our result is plotted to aid the eye. Our films were deposited at a 35$^\circ$ angle with respect to the substrate normal with assisting \emph{in-situ} magnetic field of 70~Oe at $\sim$12.8~rpm rotation speed.}
    \label{fig_deltarho}
\end{figure}

Figure \ref{fig_deltarho} shows the change in $\Delta\rho$ versus film thickness in comparison with previous studies. Mitchell \emph{et al.} \cite{Mitchell1964} and McGuire and Potter \cite{mcguire1975} claimed that $\Delta\rho$ is independent of the film thickness. While Miyazaki \emph{et al.} \cite{Miyazaki1989} concluded that there was a slight increase in $\Delta\rho$ with the film thickness and Williams \emph{et al.} \cite{Williams1986} showed a strong increase in $\Delta\rho$ with the film thickness. On the other hand Funaki \emph{et al.} \cite{Funaki1994} reported a maximum $\Delta\rho$ around 50~nm for as prepared Py which drops afterwards. However our result shows the $\Delta\rho$ increase with thickness to a maximum at 30~nm and becomes constant afterward. The sharp increase in $\Delta\rho$ between 10--30~nm seems to apear in the all data sets including Mitchell \emph{et al.} \cite{Mitchell1964} and McGuire and Potter \cite{mcguire1975}. This corresponds to the range of thicknesses when there is competition between different grains with different orientations in the case of an amorphous substrates such as SiO$_2$ \cite{Mahieu2006}. While for thicker films the dominating orientations are determined by the growth conditions and the result of different studies become more scattered.

\begin{figure}
    \centering
    \includegraphics[width=1\linewidth]{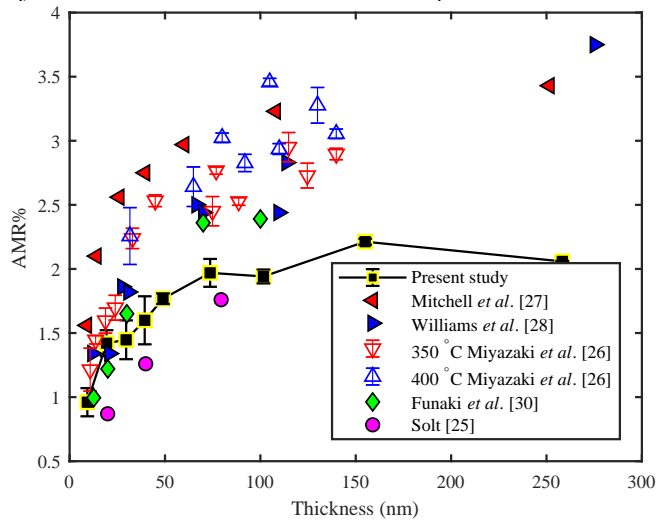}
    \caption{Variation of AMR\% with the film thickness in comparison with previous results. The solid black line in our result is plotted to aid the eye. Our films were deposited at a 35$^\circ$ angle with respect to the substrate normal with assisting \emph{in-situ} magnetic field of 70~Oe at $\sim$12.8~rpm rotation speed.}
    \label{fig_AMR}
\end{figure}

The AMR change with the thickness is shown in figure \ref{fig_AMR} which also contains previous results for comparison. McGuire and Potter \cite{mcguire1975} believed since $\Delta\rho$ is independent of the film thickness the AMR ratio only depends on $\rho_{\rm{ave}}$. Thus, a saturation behaviour with the film thickness is expected for AMR as $\rho_{\rm{ave}}$ saturates around the bulk value at higher thicknesses. While Williams \emph{et al.} \cite{Williams1986} reported an increasing trend with increase in thickness which is due to their increased $\Delta\rho$. The rest of the data sets, however, show a saturation in AMR as their $\Delta\rho$ does not grow continuously with the thickness. It is also clear that data sets with lower $\rho_{\rm{ave}}$ present higher AMR here which also indicates the importance of $\rho_{\rm{ave}}$. Thus, for the increased AMR ratio both high  $\Delta\rho$ and low $\rho_{\rm{ave}}$ is required.

\section{Conclusions}
In conclusion, it is shown that the extended vdP method is a sensitive way to obtain anisotropic resistivity in thin ferromagnetic films. This includes AMR measurements which require magnetization saturation parallel and perpendicular to the current direction. In comparison with the conventional way of using Hall-bars, the vdP method presents more consistency along the hard and easy axis of the film. Thus the AMR value measured by the extended vdP method is independent of the initial magnetization direction in the film. Also variation of microstructure and AMR with the thickness of Py films was studied. It is shown that the resistivity of the films obtained by the vdP method strongly correlates with the film thickness, grain size and texture. The AMR ratio increases at first with the film thickness and then saturates just below 100~nm thickness, as both $\Delta\rho$ and $\rho_{\rm{ave}}$ independently saturate, their ratio thus remaining fixed. The $\Delta\rho$ and $\rho_{\rm{ave}}$ seems to be affected by microstructure obtained in the preparation method.

\acknowledgements

This work was supported by the Icelandic Research Fund Grant No. 120002023.

\bibliographystyle{apsrev}
\bibliography{Ref.bib}

\end{document}